\documentclass[aps,prd,reprint,superscriptaddress]{revtex4-2}
\usepackage{subfigure}
\usepackage{tabularx}
\usepackage{graphicx}	
\usepackage[tbtags]{amsmath}	
\usepackage{amssymb}	
\usepackage{natbib}
\usepackage{color}
\usepackage{amsmath}
\usepackage{subeqnarray}
\usepackage{multirow}

\begin{document}
\title{Precision measurements of the magnetic parameters of LISA Pathfinder test masses}

%
%
%
%
%
%
%
%

\def\addressa{European Space Astronomy Centre, European Space Agency, Villanueva de la
Ca\~{n}ada, 28692 Madrid, Spain}
\def\addressb{Albert-Einstein-Institut, Max-Planck-Institut f\"ur Gravitationsphysik und Leibniz Universit\"at Hannover,
Callinstra{\ss}e 38, 30167 Hannover, Germany}
\def\addressc{APC, Univ Paris Diderot, CNRS/IN2P3, CEA/lrfu, Obs de Paris, Sorbonne Paris Cit\'e, France}
\def\addressd{High Energy Physics Group, Physics Department, Imperial College London, Blackett Laboratory, Prince Consort Road, London, SW7 2BW, UK }
\def\addresse{Dipartimento di Fisica, Universit\`a di Roma ``Tor Vergata'',  and INFN, sezione Roma Tor Vergata, I-00133 Roma, Italy}
\def\addressf{Department of Industrial Engineering, University of Trento, via Sommarive 9, 38123 Trento, 
and Trento Institute for Fundamental Physics and Application / INFN}
\def\addressh{European Space Technology Centre, European Space Agency, 
Keplerlaan 1, 2200 AG Noordwijk, The Netherlands}
\def\addressi{Dipartimento di Fisica, Universit\`a di Trento and Trento Institute for 
Fundamental Physics and Application / INFN, 38123 Povo, Trento, Italy}
\def\addressk{Istituto di Fotonica e Nanotecnologie, CNR-Fondazione Bruno Kessler, I-38123 Povo, Trento, Italy}
\def\addressj{The School of Physics and Astronomy, University of
Birmingham, Birmingham, UK}
\def\addressl{Institut f\"ur Geophysik, ETH Z\"urich, Sonneggstrasse 5, CH-8092, Z\"urich, Switzerland}
\def\addressm{The UK Astronomy Technology Centre, Royal Observatory, Edinburgh, Blackford Hill, Edinburgh, EH9 3HJ, UK}
\def\addressn{Institut de Ci\`encies de l'Espai (ICE, CSIC), Campus UAB, Carrer de Can Magrans s/n, 08193 Cerdanyola del Vall\`es, Spain}
\def\addresso{DISPEA, Universit\`a di Urbino ``Carlo Bo'', Via S. Chiara, 27 61029 Urbino/INFN, Italy}
\def\addressp{European Space Operations Centre, European Space Agency, 64293 Darmstadt, Germany}
\def\addressq{Physik Institut, 
Universit\"at Z\"urich, Winterthurerstrasse 190, CH-8057 Z\"urich, Switzerland}
\def\addressr{SUPA, Institute for Gravitational Research, School of Physics and Astronomy, University of Glasgow, Glasgow, G12 8QQ, UK}
\def\addresss{Department d'Enginyeria Electr\`onica, Universitat Polit\`ecnica de Catalunya,  08034 Barcelona, Spain}
\def\addresst{Institut d'Estudis Espacials de Catalunya (IEEC), C/ Gran Capit\`a 2-4, 08034 Barcelona, Spain}
\def\addressu{Gravitational Astrophysics Lab, NASA Goddard Space Flight Center, 8800 Greenbelt Road, Greenbelt, MD 20771 USA}
\def\addressbb{Department of Mechanical and Aerospace Engineering, MAE-A, P.O. Box 116250, University of Florida, Gainesville, Florida 32611, USA}
\def\addresscc{Istituto di Fotonica e Nanotecnologie, CNR-Fondazione Bruno Kessler, I-38123 Povo, Trento, Italy}
\def\addressdd{isardSAT SL, Marie Curie 8-14, 08042 Barcelona, Catalonia, Spain}
\def\addressee{Escuela Superior de Ingenier\'ia, Universidad de C\'adiz, 11519 C\'adiz, Spain}
\def\addressen{Observatoire de la C\^{o}te d'Azur, Boulevard de l'Observatoire CS 34229 - F 06304 NICE, France}


\author{M~Armano}\affiliation{\addressh}
\author{H~Audley}\affiliation{\addressb}
\author{J~Baird}\affiliation{\addressc}
\author{P~Binetruy}\thanks{Deceased}\affiliation{\addressc}
\author{M~Born}\affiliation{\addressb}
\author{D~Bortoluzzi}\affiliation{\addressf}
\author{E~Castelli}\affiliation{\addressi}
\author{A~Cavalleri}\affiliation{\addresscc}
\author{A~Cesarini}\affiliation{\addresso}
\author{A\,M~Cruise}\affiliation{\addressj}
\author{K~Danzmann}\affiliation{\addressb}
\author{M~de Deus Silva}\affiliation{\addressa}
\author{I~Diepholz}\affiliation{\addressb}
\author{G~Dixon}\affiliation{\addressj}
\author{R~Dolesi}\affiliation{\addressi}
\author{L~Ferraioli}\affiliation{\addressl}
\author{V~Ferroni}\affiliation{\addressi}
\author{E\,D~Fitzsimons}\affiliation{\addressm}
\author{M~Freschi}\affiliation{\addressa}
\author{L~Gesa}\thanks{Deceased}\affiliation{\addressn}\affiliation{\addresst}
\author{D~Giardini}\affiliation{\addressl}
\author{F~Gibert}\affiliation{\addressi}\affiliation{\addressdd}
\author{R~Giusteri}\affiliation{\addressb}
\author{C~Grimani}\affiliation{\addresso}
\author{J~Grzymisch}\affiliation{\addressh}
\author{I~Harrison}\affiliation{\addressp}
\author{M-S~Hartig}\affiliation{\addressb}
\author{G~Heinzel}\affiliation{\addressb}
\author{M~Hewitson}\affiliation{\addressb}
\author{D~Hollington}\affiliation{\addressd}
\author{D~Hoyland}\affiliation{\addressj}
\author{M~Hueller}\affiliation{\addressi}
\author{H~Inchausp\'e}\affiliation{\addressc}\affiliation{\addressbb}
\author{O~Jennrich}\affiliation{\addressh}
\author{P~Jetzer}\affiliation{\addressq}
\author{N~Karnesis}\affiliation{\addressc}
\author{B~Kaune}\affiliation{\addressb}
\author{N~Korsakova}\affiliation{\addressen}
\author{C\,J~Killow}\affiliation{\addressr}
\author{L~Liu}\affiliation{\addressi}
\author{J\,A~Lobo}\thanks{Deceased}\affiliation{\addressn}\affiliation{\addresst}
\author{J\,P~L\'opez-Zaragoza}\email{jplopez@ice.csic.es}\affiliation{\addressn}\affiliation{\addresst}
\author{R~Maarschalkerweerd}\affiliation{\addressp}
\author{D~Mance}\affiliation{\addressl}
\author{V~Mart\'{i}n}\affiliation{\addressn}\affiliation{\addresst}
\author{J~Martino}\affiliation{\addressc}
\author{L~Martin-Polo}\affiliation{\addressa}
\author{F~Martin-Porqueras}\affiliation{\addressa}
\author{N~Meshksar}\affiliation{\addressl}
\author{P\,W~McNamara}\affiliation{\addressh}
\author{J~Mendes}\affiliation{\addressp}
\author{L~Mendes}\affiliation{\addressa}
\author{M~Nofrarias}\email{nofrarias@ice.csic.es}\affiliation{\addressn}\affiliation{\addresst}
\author{S~Paczkowski}\affiliation{\addressb}
\author{M~Perreur-Lloyd}\affiliation{\addressr}
\author{A~Petiteau}\affiliation{\addressc}
\author{P~Pivato}\affiliation{\addressi}
\author{E~Plagnol}\affiliation{\addressc}
\author{J~Ramos-Castro}\affiliation{\addresss}\affiliation{\addresst}
\author{J~Reiche}\affiliation{\addressb}
\author{D\,I~Robertson}\affiliation{\addressr}
\author{F~Rivas}\affiliation{\addressn}\affiliation{\addresst}
\author{G~Russano}\affiliation{\addressi}
\author{L~Sala}\affiliation{\addressi}
\author{D~Serrano}\email{dserrano@ice.csic.es}\affiliation{\addressn}\affiliation{\addresst}
\author{J~Slutsky}\affiliation{\addressu}
\author{C\,F~Sopuerta}\affiliation{\addressn}\affiliation{\addresst}
\author{T~Sumner}\affiliation{\addressd}
\author{D~Texier}\affiliation{\addressa}
\author{J\,I~Thorpe}\affiliation{\addressu}
\author{D~Vetrugno}\affiliation{\addressi}
\author{S~Vitale}\affiliation{\addressi}
\author{G~Wanner}\affiliation{\addressb}
\author{H~Ward}\affiliation{\addressr}
\author{P\,J~Wass}\affiliation{\addressd}\affiliation{\addressbb}
\author{W\,J~Weber}\affiliation{\addressi}
\author{L~Wissel}\affiliation{\addressb}
\author{A~Wittchen}\affiliation{\addressb}
\author{P~Zweifel}\affiliation{\addressl}

%
%
%
%
%
%
%

%

\begin{abstract}

A precise characterization of the magnetic properties of LISA Pathfinder free falling test-masses is of special interest for future gravitational wave observatory in space. Magnetic forces have an important impact on the instrument sensitivity in the low frequency regime below the millihertz. In this paper we report on the magnetic injection experiments performed throughout LISA Pathfinder operations. We show how these experiments allowed a high precision estimate of the instrument magnetic parameters. The remanent magnetic moment was found to have a modulus of $(0.245\pm0.081)\,\rm{nAm}^2$, the x-component of the background magnetic field within the test masses position was measured to be $(414 \pm 74)$ nT and its gradient had a value of $\rm(-7.4\pm2.1)\,\mu$T/m. Finally, we also measured the test mass magnetic susceptibility at 5 mHz to be $(-3.3723\pm0.0069)\times$10$^{-5}$. All results are in agreement with on-ground estimates.

\end{abstract}

\maketitle


\section{Introduction} \label{sec:intro}

LISA Pathfinder (LPF)~\citep{Anza05, Antonucci12} was an ESA mission designed as a technology demonstrator for the future gravitational wave observatory in space, LISA~\citep{Amaro17}. The main goal of the mission was to demonstrate key technologies required to detect gravitational waves in space.
In order to do so, the instrument on-board had to demonstrate a relative acceleration noise between its two test masses (TMs) in nominal geodesic motion at a level of $\rm 3 \times 10^{-14}\,ms^{-2}Hz^{-1/2}$ at 1 mHz, a level of precision impossible to achieve with on ground gravitational wave detectors.

LPF launched on December $\rm 3^{rd}$, 2015 and started its scientific operations on the March $\rm 1^{st}$, 2016 after reaching the Lagrange point L1 of the Earth-Sun system. The mission was divided into two different experiments on-board, the European Space Agency LISA Technology Package (LTP) and the NASA Disturbance Reduction System (DRS). After seventeen months of scientific operations, the mission successfully demonstrated its main scientific goal, surpassing its requirements and achieving a level of acceleration noise below the LISA requirements in its entire measurement frequency band~\citep{Armano16, Armano18}.

As important as achieving this demanding level of geodesic free fall was the development of an understanding all the different contributions that build the noise model of the instrument. Several experiments were planned during the LPF operations in order to isolate and evaluate the most important contributions to the acceleration noise budget. With that objective, LISA Pathfinder carried the Data and Diagnostics Subsystem (DDS), which included a temperature measurement subsystem~\citep{Sanjuan07, Armano19_Temp}, a magnetic diagnostic subsystem~\citep{DiazAguilo13, Armano20_Mag} and a radiation monitor~\citep{Canizares09, Canizares11, Armano18_GCR1, Armano18_GCR2}. 

In this work we will focus on the results of the magnetic diagnostics and, specifically, on the experiments run to characterize  the magnetic parameters of the test masses on-board LPF. Precise knowledge of such values is crucial for the future space-borne gravitational wave observatories, since any magnetic perturbation can have a potential impact on the instrument performance through magnetic parasitic forces.

This work is organized as follows. In section \ref{sec:setup} we describe the magnetic diagnostic system on-board designed to study and disentangle the nature of the magnetic forces,  introduced in section \ref{sec:forces}, that can perturb the test mass motion. In section \ref{sec:experiments} we describe the in-flight magnetic experiments performed to extract the TMs magnetic parameters and we present our conclusions in section \ref{sec:conclusions}.

\section{Experimental setup} \label{sec:setup}

\subsection{The magnetic diagnostics subsystem}

The magnetic diagnostics subsystem on-board LISA Pathfinder was responsible for monitoring the magnetic environment and creating controlled magnetic fields to perturb the test mass motion in order to properly characterize the contribution of magnetic forces to the total instrument noise budget. To achieve these goals, the subsystem was composed by four triaxial magnetometers and two induction coils.

\begin{figure}[t]
\begin{center}
\includegraphics[width=0.6\textwidth]{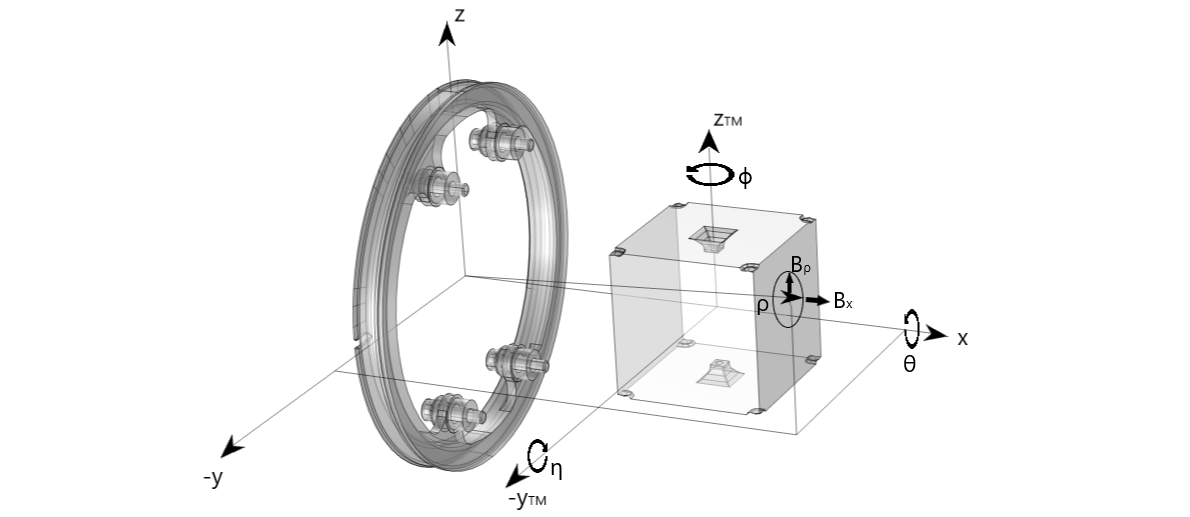} 
\caption{Coordinate reference system for the coil and the test mass. The convention used for the three angles of rotations along each test mass axis is also shown.}
\label{fig.coilsScheme}
\end{center}
\end{figure}

The coils---see Fig.~\ref{fig.coilsScheme}---were able to produce a controlled magnetic field at the TM locations as well as in the magnetometers closest to them. Both circular induction coils, with an average radius of 56.5 mm, were located 85.5 mm away from the test masses and they were attached to the external wall of each vacuum enclosure. The wire winding used to build up the coils were made of a Titanium alloy (Ti$_{6}$Al$_{4}$V) and loop around the structure for a total of 2400 turns. The centers of both coils were aligned with the axis, $x$, joining both TMs centers so that the induced magnetic field had axial symmetry. The four magnetometers were aligned by pairs in the $x-y$ plane of the TMs in order to be able to measure gradients within the spacecraft in both the $x$ and $y$ directions. Magnetic field gradients along the $z$ direction could not be measured.
The magnetometers continuously measured the evolution of the on-board magnetic field with a precision of 10\,nT\,Hz$^{-1/2}$. Each fluxgate magnetometer measurement axis consisted of a sensing coil surrounding a second inner drive coil around a high permeability magnetic core material. This meant that all four magnetometers contained active magnetic sensors that had to be located far enough from the test masses for them not to contribute as a source of magnetic parasitic forces.

\subsection{Magnetic environment on-board} 

The background magnetic field measured on-board was completely dominated by the contribution from the electronics of the spacecraft units. Among them, the thruster systems were a major contributor, both the cold gas high pressure latch valves (the ones used by ESA) and the colloidal thrusters (the ones operated by NASA). Cold gas thrusters or, more precisely, some permanent magnets in the cold gas thruster subsystem,  contributed with roughly the 80\% of the measured magnetic field. Although a strong contribution, this remained constant throughout the mission---partially thanks to the high thermal stability reached on-board~\citep{Armano19_Temp}. This was not the case for the colloidal thrusters, where a persistent slow drift of around 150\,nT in the span of 100 days was observed~\citep{Armano20_Mag}.
The main contribution of the magnetic-induced force noise is below the millihertz~\citep{Armano23_Mag_PRL}. In this frequency regime, magnetic field fluctuations are dominated by the interplanetary magnetic field contribution, which can show an important non-stationary component associated with changes in the interplanetary plasma. For instance, variations in the range of $\rm 300-500\, km\, s^{-1}$ in the solar wind velocity were found to be correlated to variations in the magnetic fields amplitude spectral density in the range $\rm 20-50\, \mu Hz$ of around $\rm 170-750 \,nT \ Hz^{-1/2}$~\citep{Armano20_Mag}.


In what refers to our analysis in the following, we assume that in all the in-flight experiments where we induced magnetic fields with the coils, the background magnetic field (either generated by the spacecraft or due to the interplanetary contribution) can be safely neglected as it was at least one order of magnitude smaller than the ones induced by the coils. The same is true for the gradients of the magnetic fields.

\subsection{Forces on-board LPF}

We evaluate the induced force in the test mass through the $\Delta g$ variable---the principal scientific output of the mission---nominally defined as the differential acceleration between the two TMs in their nominal position~\cite{Armano16}. Since the main objective of the $\Delta g$ measurement is the evaluation of the free fall of the test masses, those forces arising due to the spacecraft dynamics control loop or other forces caused by spacecraft non-inertial reference frame are subtracted in the definition of this parameter~\cite{Armano18_xtalk}. We will assume that during magnetic injections the dominant forces that TMs will feel will be purely of magnetic origin. $\Delta g$, by construction, is the difference in acceleration between TMs along the axis joining them, the same as the spacecraft $x$ axis by definition and aligned with the $x$ axis from Fig.~\ref{fig.coilsScheme}, 
\begin{equation}
\Delta g = \dfrac{ F_{2, x} }{ M_{\rm{TM1}} } - \dfrac{ F_{1, x} }{ M_{\rm{TM2}} },
\label{eq.deltaG}
\end{equation}
where $M_{\rm{TM}}$ is the mass of each TM and $F$ is the total force each one feels. When calculating the force measured by each TM during magnetic injections, we assume that the TM furthest from the injecting coil will feel a negligible force when compared to the one nearest to the coil. So, when injecting magnetic signals with the coils, we will estimate the magnetic force that the nearest TM feels as $F_{x} \equiv \rm{M_{TM}}\Delta g$. 

The same is true for the torque such that $N_{\eta,\phi} \equiv I\Delta g_{\eta, \phi}$, where $I$ is the moment of inertia of a cube and $\Delta g_{\eta, \phi}$ are the angular accelerations measured by the interferometric system. Only rotations along the $y$ and $z$ axes, $\eta$ and $\phi$ respectively in Fig~\ref{fig.coilsScheme}, are obtainable since the instrument is not sensitive to rotations around the $x$ axis as it is aligned with the laser beam. 

\section{Magnetic-induced forces and torques in a free falling test mass} \label{sec:forces}

Magnetic fluctuations can couple into the dynamics of the free falling test masses on-board the satellite. In the following we develop the basic equations needed to describe the experiments carried out with the induction coils in LISA Pathfinder.

\subsection{A magnetic dipole in a surrounding magnetic field}

In a first approximation, the free-falling test masses inside LISA Pathfinder 
can be considered as a magnetic dipole with total magnetic moment density $\mathbf{m}$ inside of a 
surrounding magnetic field $\mathbf{B}$. Parameters in bold refer to vectors. The dipole would therefore feel an associated force and torque given by
\begin{subequations}
\label{eq.forceTorque}
\begin{align}
\mathbf{F} & =  \left\langle \left(\mathbf{m} \cdot \mathbf{\nabla} \right)\mathbf{B} \right\rangle \rm{V}, \label{eq.Forceone} \\
\mathbf{N}  & =  \left\langle \mathbf{m} \times \mathbf{B} + \mathbf{r} \times (\mathbf{m} \cdot \mathbf{\nabla})\mathbf{B} \right\rangle \rm{V},\label{eq.torquone}
\end{align}
\end{subequations}
where $\mathbf{r}$ denotes the distance to the TM with respect to the coil. We use the convention $\left\langle \ldots\ \right\rangle \equiv \dfrac{1}{\rm{V}}\int_{\rm{V}} \,(\ldots)\,d^{3}x$ to denote the average of the enclosed quantity over the TM volume, V. The total magnetic moment density $\mathbf{m}$ is the sum of two components:
the remanent magnetic moment density $\mathbf{m_r}$ which depends on the material and manufacturing process 
and the induced magnetic moment density, $\mathbf{m_i}$. The induced magnetic dipole density of any material is proportional to the applied magnetic field, i.e. 
\begin{equation}
\mathbf{m_i} = \chi / \mu_o \, \mathbf{B},
\label{eq.inducedMoment}
\end{equation}
where $\chi$ is the magnetic susceptibility and $\mu_{0}$ is the vacuum permeability. The test mass composition is 73\%\,gold, diamagnetic, and 27\%\,platinum, paramagnetic. Given that the dominant material is gold we should expect the TM to behave like any diamagnetic material, opposing the external magnetic field thus, having a negative magnetic susceptibility. Despite this, we leave the sign undetermined in the following derivation. Also, we have implicitly assumed here an isotropic test mass which allows the use of a scalar susceptibility in the previous equation. Next, we assume the test mass magnetized by a slowly oscillating field 
\begin{equation}
\mathbf{B}(t) = \mathbf{B}^{AC} \sin(\omega \, t),
\end{equation}
where $\mathbf{B}^{AC}$ is the amplitude of the oscillating magnetic field and $\omega$ its angular frequency, 
we will obtain a magnetization that varies with time accordingly, $\mathbf{m_{i}}(t)$. 
In diamagnetic, paramagnetic and many ferromagnetic materials, the magnetization also varies 
sinusoidally and in phase with the applied magnetic field with a constant ratio given by the magnetic susceptibility. However, some ferromagnetic materials show a delayed response that is not in phase with the applied field. 
This phenomena is typically described by considering the in-phase or real, $\chi_{r}$, and the out-of-phase or imaginary, $\chi_{i}$, components 
of the magnetic susceptibility such that $\chi = \chi_{r} + i\chi_{i}$, where $i=\sqrt{-1}$. For the case of LPF experiments the most relevant physical mechanism involved in the imaginary susceptibility are eddy currents since this contribution becomes increasingly important with increasing conductivity of the material. However, for most of the experiments in the low frequency regime this component is expected to be orders of magnitude smaller than the real part thus, its contribution can be neglected. In the following, therefore, we refer to the magnetic susceptibility $\chi$ as equivalent to its real component $\chi_{r}$, unless explicitly stated otherwise.

Considering both contributions of the magnetization (the remanent and the induced magnetic moments) we expand Eqs.~\eqref{eq.forceTorque} into
\begin{subequations}
\begin{align}
\mathbf{F} & = \left\langle \left( \mathbf{m_r} \cdot \mathbf{\nabla}\right)\mathbf{B} + \frac{\chi}{\mu_0}\,\left[\left(\mathbf{B} \cdot \mathbf{\nabla} \right)\mathbf{B} \right] \right\rangle\rm{V},
\label{eq.Force} \\
\mathbf{N} & =  
\left\langle  
\mathbf{m_r} \times \mathbf{B} 
+ \mathbf{r}\times \left[ (\mathbf{m_r} \cdot \nabla )\, \mathbf{B} 
+ \frac{\chi}{\mu_0} (\mathbf{B} \cdot \mathbf{\nabla})\, \mathbf{B}     
\right] 
\right\rangle\rm{V}. 
\label{eq.Torque}
\end{align}
\end{subequations}

In order to describe our experiments in the following sections, 
we need to develop the equations further.
First, we will consider the magnetic field as composed by an applied, oscillating magnetic field 
$\mathbf{B}^{AC}$, and a stable magnetic field $\mathbf{B}_{0}$, divided into an applied time independent DC magnetic field $\mathbf{B}^{DC}$ and some environmental background $\mathbf{B}_{back.}$
\begin{equation}
\begin{split}
\mathbf{B}  &= \mathbf{B}_0 +  \mathbf{B}^{AC} \sin(\omega \, t) \\&= \left(\mathbf{B}_{back.} + \mathbf{B}^{DC}\right) +  \mathbf{B}^{AC} \sin(\omega \, t).
\label{eq.magfield}
\end{split}
\end{equation}
By substituting in Eq.~\eqref{eq.Force} and factoring out the components in terms of their frequency response to the input signal, we find that the force can be divided into three components: a constant DC term, a term that oscillates at the same frequency of the induced magnetic field  $1\, \omega$ and a term oscillating at twice the frequency $2\, \omega$
\begin{equation}
 \mathbf{F} = \mathbf{F_{DC}} + \mathbf{F_{1\omega}} + \mathbf{F_{2\omega}},
 \label{eq.forces}
\end{equation}
with
\begin{subequations}
\label{eq.3forces}
\begin{align}
\begin{split}
\mathbf{F_{DC}} =  \Biggl[& \left\langle \left(\mathbf{M_r} \cdot \mathbf{\nabla}\right) \mathbf{B}_0  \right\rangle
\\&+ \frac{\chi\rm{V}}{\mu_{0}} \left( \left\langle \left(\mathbf{B_{0}} \cdot \mathbf{\nabla}\right) \mathbf{B_{0}} \right\rangle 
+ \frac{1}{2} \left\langle (\mathbf{B}^{AC} \cdot \mathbf{\nabla}) \mathbf{B}^{AC} \right\rangle \right) 
\Biggr],
\label{eq.force0w}\\
\end{split}
\\[2ex]
\begin{split}
\mathbf{F_{1\omega}} =  \Biggl[& \left\langle\left( \mathbf{M_r} \cdot \mathbf{\nabla}\right) \mathbf{B}^{AC}  \right\rangle
\\&+ \frac{\chi\rm{V}}{\mu_0} \biggl( \left\langle \left(\mathbf{B}_0 \cdot \mathbf{\nabla}\right) \mathbf{B}^{AC} \right\rangle + \left\langle \left(\mathbf{B}^{AC} \cdot \mathbf{\nabla}\right) \mathbf{B}_0 \right\rangle \biggr) 
\Biggr]\\& \times\sin (\omega\, t),
\label{eq.force1w}\\
\end{split}
\\[0ex]
\begin{split}
\mathbf{F_{2\omega}} =  \left[ -\frac{\chi\rm{V}}{2\mu_{0}} \left\langle \left(\mathbf{B}^{AC} \cdot \mathbf{\nabla}\right) \mathbf{B}^{AC} \right\rangle \right]\cos (2\, \omega\, t),
\label{eq.force2w}
\end{split}
\end{align}
\end{subequations}
where $\mathbf{M_r} = \mathbf{m_r}\rm{V}$ is the remanent magnetic moment and we have assumed homogeneity and stationarity of the test mass properties. Considering that the relative acceleration measurements in LISA Pathfinder are in the $x$ direction, the only component of the force from Eq.~\ref{eq.forces} that will be needed is its $x$ component. Analogously, if we manipulate the torque equations a similar result with the three terms before mentioned should appear.

\subsection{Estimate of test mass magnetic parameters}\label{sec:extraction}

The evaluation of both the force and torque expressions, Eqs.~\eqref{eq.Force} and \eqref{eq.Torque} respectively, implies the calculation of the average of an external magnetic field and its gradient within the TMs volume as expressed by $\left\langle \ldots\ \right\rangle$. Making use of the induction coils from Fig.~\ref{fig.coilsScheme} we can control the injected field ($\mathbf{B^{AC,\,DC}}$ and $\mathbf{\nabla B^{AC,\,DC}}$) as it can be calculated by means of Ampère's induction laws under the assumptions of coils with negligible thickness and a wire winding of N turns. Thanks to the symmetry of our system, only the $x$ components of the averaged induced fields are non-zero, $\left\langle B_{x}^{AC,\,DC}\right\rangle$, and their gradients along the $y$ and $z$ axes are 3 orders of magnitude smaller than $\left\langle\nabla_{x}B_{x}^{AC,\,DC}\right\rangle$. Furthermore, the magnetic field in the $x$ direction and its gradient along $x$ are proportional to one another at any given point in space, that is: $\left\langle B_{x}^{AC,\,DC}\right\rangle = \kappa\left\langle\nabla_{x}B_{x}^{AC,\,DC}\right\rangle$. This factor constant $\kappa$ only depends on the coil dimensions and the distance from the coil center. Its value can be found analytically for the simple on-axis magnetic field of a coil but it is harder to obtain for the general off-axis magnetic field formula involving elliptic integrals. Thus, its value was calculated numerically to be $\kappa = -0.04487$ m for our particular configuration, with negligible uncertainty originated only due to numerical error. We refer the interested reader to Appendix~\ref{sec:annex} for more detail on the calculations involved at the TMs location. Finally, the magnetic force is obtained by applying a heterodyne demodulation at the different frequencies of interest of the on-board measurements of the stray TM force (more detail on this in the upcoming section) resulting in the estimators $\hat{F}_{DC,x}$ , $\hat{F}_{1\omega,x}$ and $\hat{F}_{2\omega,x}$. We now proceed to describe how we will estimate the test mass magnetic parameters from the previous generic expressions.

\paragraph{Magnetic susceptibility}

The coupling between an induced magnetic field and its gradient with the magnetic susceptibility of the test mass is responsible for the appearance of a force component at twice the injected modulation frequency in Eq.~\eqref{eq.force2w}. Our analysis can take advantage of this by extracting the signal at $2\omega$ from the measured test mass force, i.e.
\begin{equation}
\chi_{2\omega} = 
- \frac{2 \, \mu_0}{V} 
\frac{\hat{F}_{2\omega,x}}{\left\langle B_{x}^{AC}\right\rangle \cdot \left\langle\nabla_{x} B_{x}^{AC}\right\rangle}.
\label{eq.susc_estimate}
\end{equation}
This equation provides a direct estimate of the test mass susceptibility decoupled from any other of the magnetic parameters.
The notation in Eq.~\eqref{eq.susc_estimate} shows explicitly that the estimate of the susceptibility is obtained at twice the injected frequency by demodulating the encoded information in $\Delta g$ and comparing it with the predicted TM average magnetic field and gradient.

We notice that, in principle, we could use the signal at 2$\omega$ to obtain both real and imaginary contributions to the magnetic susceptibility.
To estimate the imaginary contribution we would need to look for a 2$\omega$ contribution with a $\pi/2$ phase shift with respect the original injection. We will explore this in the discussion of our results in Section~\ref{sec:experiments}.

\paragraph{Remanent magnetic moment} \label{f1omegan1omega}

The component of the force at the injection frequency, $F_{1\omega,x}$, depends on all the magnetic parameters of the TM. Taking advantage of the fact that all terms in Eq.~\eqref{eq.force1w} depend on $\left\langle B_{x}^{AC}\right\rangle$ or $\left\langle\nabla_{x}B_{x}^{AC}\right\rangle$, we can rewrite the expression as follows
\begin{equation}
\hat{F}_{1\omega,x}  =  \left[ M_{r,x} + \frac{\chi V}{\mu_0} ( B_{0,x} + \kappa \nabla_{x} B_{0,x}) \right] \left\langle\nabla_{x} B_{x}^{AC}\right\rangle.
\label{eq.F1w_ACs}
\end{equation}
The term in brackets can be related to an effective magnetic moment such that
\begin{equation}
\begin{split}
M_{\text{eff}f,x}  =  M_{r,x}+ \frac{\chi V}{\mu_0} \Biggl[ & \left(B_{back.,x} + \left\langle B_{x}^{DC}\right\rangle\right) \\& + \kappa \nabla_{x} \left(B_{back.,x} + \left\langle B_{x}^{DC}\right\rangle \right) \Biggr],
\label{eq.Meffec}
\end{split}
\end{equation}
where we have expanded $B_{0,x}$ as explained in Eq.~\eqref{eq.magfield}. $B_{back.,x}$ can be considered negligible compared to the injected magnetic field $\left\langle B_{x}^{DC}\right\rangle$ as its value is expected to be an order of magnitude smaller than the injected fields through the coils. Thus, we have
\begin{equation}
M_{\text{eff},x} \simeq
 M_{r,x} + \frac{2\, \chi V}{\mu_0} \left\langle B_{x}^{DC}\right\rangle.
\label{eq.Meffapprox}
\end{equation}

If the only variable in Eq.~\eqref{eq.Meffapprox} is $\left\langle B_{x}^{DC}\right\rangle$, then plotting Eq. (\ref{eq.F1w_ACs}), we will obtain a straight line with an offset that corresponds to the remanent magnetic moment $M_{r,x}$ and a slope that is proportional to the magnetic susceptibility $\chi$ at $1\omega$. Furthermore, when we induce a magnetic field in the TM position, using the coils, apart from direct forces in the $x$ direction, we are also generating torques, as described in Eq.~\eqref{eq.Torque}. Due to the symmetry of the system the term involving the cross product with $\mathbf{r}$ will integrate to zero accross the TM volume due to the alignment between the coil axis and the TMs center resulting in only two components, see Appendix~\ref{sec:torque} for details, where only the $1\omega$ term will be of interest leading to the following equations
\begin{equation}
 \hat{N}_{\phi,1\omega}= -M_{r,y} \left\langle B_{x}^{AC} \right\rangle \ \ ; \ \ \hat{N}_{\eta,1\omega}=M_{r,z}\left\langle B_{x}^{AC}\right\rangle.
\label{eq.fortorques}
\end{equation}

We can conclude that the $1\omega$ oscillation of the torque in $\phi$ is directly related to the remanent magnetic moment along $y$, $M_{r,y}$, while the $1\omega$ oscillation of the torque in $\eta$ is directly related to the remanent magnetic moment along $z$, $M_{r,z}$. Therefore, by demodulating the torque at $1\omega$ for $\eta$ and $\phi$ and considering the values of the injected magnetic field $\left\langle B_{x}^{AC}\right\rangle$, we will be able to determine $M_{r,y}$ and $M_{r,z}$.

\paragraph{Background estimates}

Similarly to the $1\omega$ term, in Eq.~\eqref{eq.force0w}, the expression of the DC force component of the signal involves again all the unknown parameters. If we group all the terms of Eq.~\eqref{eq.force0w} as a function of $\left\langle B_{x}^{DC}\right\rangle$, we can rewrite it as
\begin{align}
\begin{split}
&\hat{F}_{DC, x}  \simeq \left(\frac{\chi V}{\mu_0 \kappa} \right)\left\langle B_{x}^{DC}\right\rangle^{2}
\nonumber\\
\end{split}
\\[0ex]
\begin{split}
+ \left[ \frac{M_{r,x}}{\kappa} + \frac{\chi V}{\mu_0} \left(\nabla_{x} B_{back., x} + \frac{B_{back., x}}{\kappa}\right) \right]\left\langle B_{x}^{DC}\right\rangle
\nonumber\\
\end{split}
\end{align}
\begin{align}
\begin{split}
+ \Biggl\{  M_{+} \nabla_{x}B_{back., x} + \frac{\chi V}{\mu_0} \Biggl[& 3B_{back., x}\nabla_{x}B_{back., x} \\&+ \dfrac{1}{2}\left\langle B_{x}^{AC}\right\rangle\left\langle\nabla_{x}B_{x}^{AC}\right\rangle \Biggr]\Biggr\},
\label{eq.forceDC_DCs}
\end{split}
\end{align}
where $M_{+} = M_{r,x} + M_{r,y} + M_{r,z}$. We have made the assumption that the background magnetic field is the same in all directions, $B_{back., x}  \simeq B_{back., y}  \simeq B_{back., z}$, because we don't have any \emph{a priori} information about its value. We also assume that the three components of $\nabla B_{back., x}$ are equal: $\nabla_{x}B_{back.,x}  \simeq \nabla_{y}B_{back.,x}  \simeq \nabla_{z}B_{back.,x}$, which is a worst case scenario since all components contributing to the background gradient would add up when in reality they could partially cancel each other. If the only variable is $\left\langle B_{x}^{DC}\right\rangle$, we can observe that $\hat{F}_{DC,x}$ has a quadratic form.

\section{In-flight experimental campaign}\label{sec:experiments}

Soon after LPF started scientific operations, on March 1st, 2016, magnetic experiments were scheduled to extract the magnetic parameters related to the TMs. The experiments consisted in applying an electric current through the coils to induce a magnetic field in the position of the TMs. The applied current in the coils was a sinusoidal signal $I(t) = I^{DC} + I^{AC} \sin ( \omega\, t )$,  where $I^{DC}$ was a constant offset, $I^{AC}$ the amplitude of the sinusoidal signal and $\omega$ its angular frequency. The current induces a magnetic field in the surroundings of the coil of the same type $B(t) = B^{DC} + B^{AC} \sin(\omega t)$.

\begin{figure}[t]
\includegraphics[width=\columnwidth]{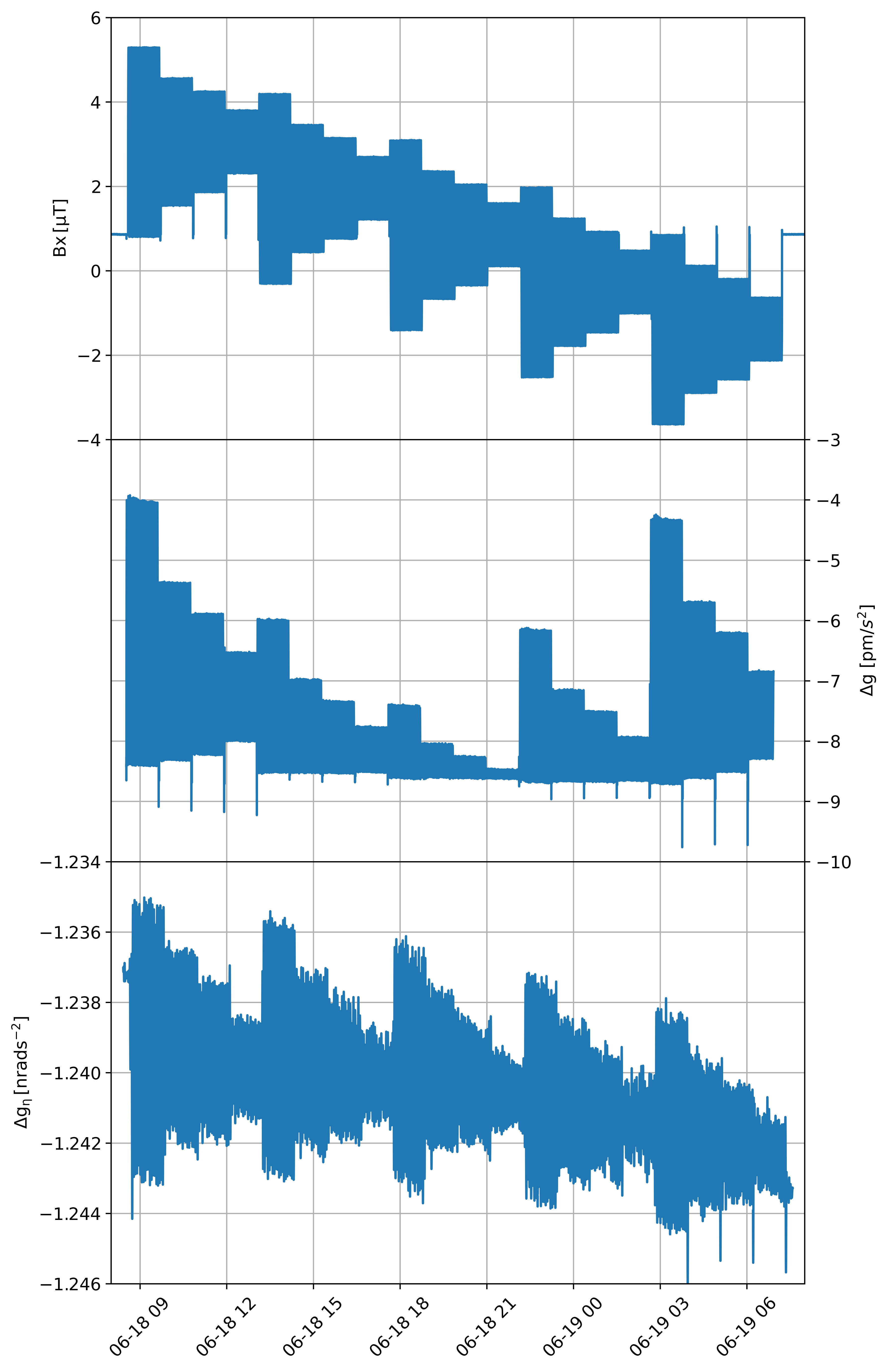}
\caption{Experiments with coil \#1, June $18^{th}$, 2016. \emph{Top}: $B_{x}$ as measured in the magnetometer closest to the coil, PX.  \emph{Middle}: $\Delta g$.  \emph{Bottom}: Angular acceleration along the rotation angle $\eta$.}
\label{fig:InjectionsJune}
\end{figure}

At the beginning of the commissioning period, all subsystems went through an initial checkout procedure. In this initial phase, coil $\#2$ ---the one closest to TM2--- showed a malfunction. Due to this fact, the injections performed during the operations period were on coil $\#1$, and the only set of injections performed in coil $\#2$ were done with low currents to prevent any possible current leak to other systems. This resulted in a reduction of the precision achievable with  coil $\#2$ experiments. Thus, most of the results that will be shown here will be for TM1 if not specified otherwise.

We carried out a total of three sets of magnetics injections. The first set was performed on the April $\rm 28^{th}$ and  $\rm 29^{th}$, 2016. The injections of April $\rm 28^{th}$ consisted of applying a sinusoidal signal through the coil $\#1$ with different DC offsets and different AC amplitudes. The injections of April $\rm 29^{th}$ were exactly the same but through coil $\#2$. The second set of injections were done on June $\rm 18^{th}$, 2016. They consisted of a series of sinusoidal injections at a wider range of both DC offsets and AC amplitudes than the previous ones but performed exclusively in coil $\#1$. The third, and last, set of injections were performed on March $\rm 14^{th}$,  $\rm 15^{th}$ and  $\rm 16^{th}$ , 2017. They consisted of very long-lasting signals at high DC offsets and with small AC amplitudes applied exclusively through coil $\#1$. The complete list of experiments is shown in Appendix~\ref{sec.App_runs}.

A typical run of magnetic experiments is shown in Fig.~\ref{fig:InjectionsJune}, where we show the results from the third set of injections during June $\rm 18^{th}$. The applied currents through the coil at 5 mHz can be seen in the Appendix within Table~\ref{tab.appendix_June}. The three panels display the main variables of interest in our analysis, these are the magnetic field in the x direction, as measured by the closest magnetometer to each coil, the acceleration produced between the TMs due to the presence of these injections and the torque being induced between both TMs along the $y$ axis.

The estimation of magnetic forces experienced by the test masses during the magnetic injections is performed by demodulating the $\Delta g$ measurements. Terms $\hat{F}_{DC,x}$, $\hat{F}_{1\omega,x}$ and $\hat{F}_{2\omega,x}$ previously defined in Section~\ref{sec:extraction} are estimated by applying a heterodyne demodulation at the corresponding frequencies and rescaling the amplitudes obtained by means of the mass of the TMs, $M_{\rm{TM}} = (1.928 \pm 0.001)$ kg. Analogously, the same procedure can be extrapolated to the torque by using the moment of inertia of a cube with the side length of the TMs, $l_{\rm{TM}} = \rm (46.000 \pm 0.005)$ mm.

Since the magnetic field can not be directly measured in the test mass position we must refer to the magnetometers read-out for calibration. We measured the amplitude in the $+x$ magnetometer of each of the 20 injections of June $\rm 18^{th}$, 2016 for coil $\#1$. To do so we demodulate the read-out at the injection frequency. By comparing these amplitudes to the ones predicted by Ampère's law with origin at the coil location we found a systematic discrepancy of $11.85 \pm 0.45 \%$, with the predicted magnetic field being larger than the measurements from the magnetometers. This systematic discrepancy can have many origins. For the magnetic field prediction, we are assuming a perfect alignment of both the coil and magnetometer, as well as no tilts. In reality, during both manufacturing and integration of the magnetic diagnostics items (coils and magnetometers) there are unavoidable mechanical tolerances. Furthermore, the launch itself adds more uncertainty on the relative distance and tilts between the diagnostics items, being this a systematic error that is difficult to quantify a priori. All the effects combined can add up to the value reported. Since the magnetometer accuracy is 0.5\%, we assume that the mismatch originates in the generation of the magnetic field by the coils and we apply this correction to all the calculated magnetic field values that intervene in all the equations derived in the previous section, redefining them for simplicity, i.e. $B_{x}^{AC} \equiv 0.8815\left\langle B_{x}^{AC}\right\rangle$.

\subsection{Remanent magnetic moment} \label{Remanent}

The estimate of the remanent magnetic moment is obtained through the dependence with the $1\omega$ component of the force expressed in Eq.~\eqref{eq.F1w_ACs} together with the approximation of Eq.~\eqref{eq.Meffec}, that we recall here for convenience:

\begin{equation}
\hat{F}_{1\omega,x}  =  \left[ M_{r,x} + \frac{2\chi V}{\mu_0} B_{x}^{DC} \right]\nabla_{x} B_{x}^{AC}. \nonumber
\end{equation}

\begin{figure}[t]
\includegraphics[width=\columnwidth]{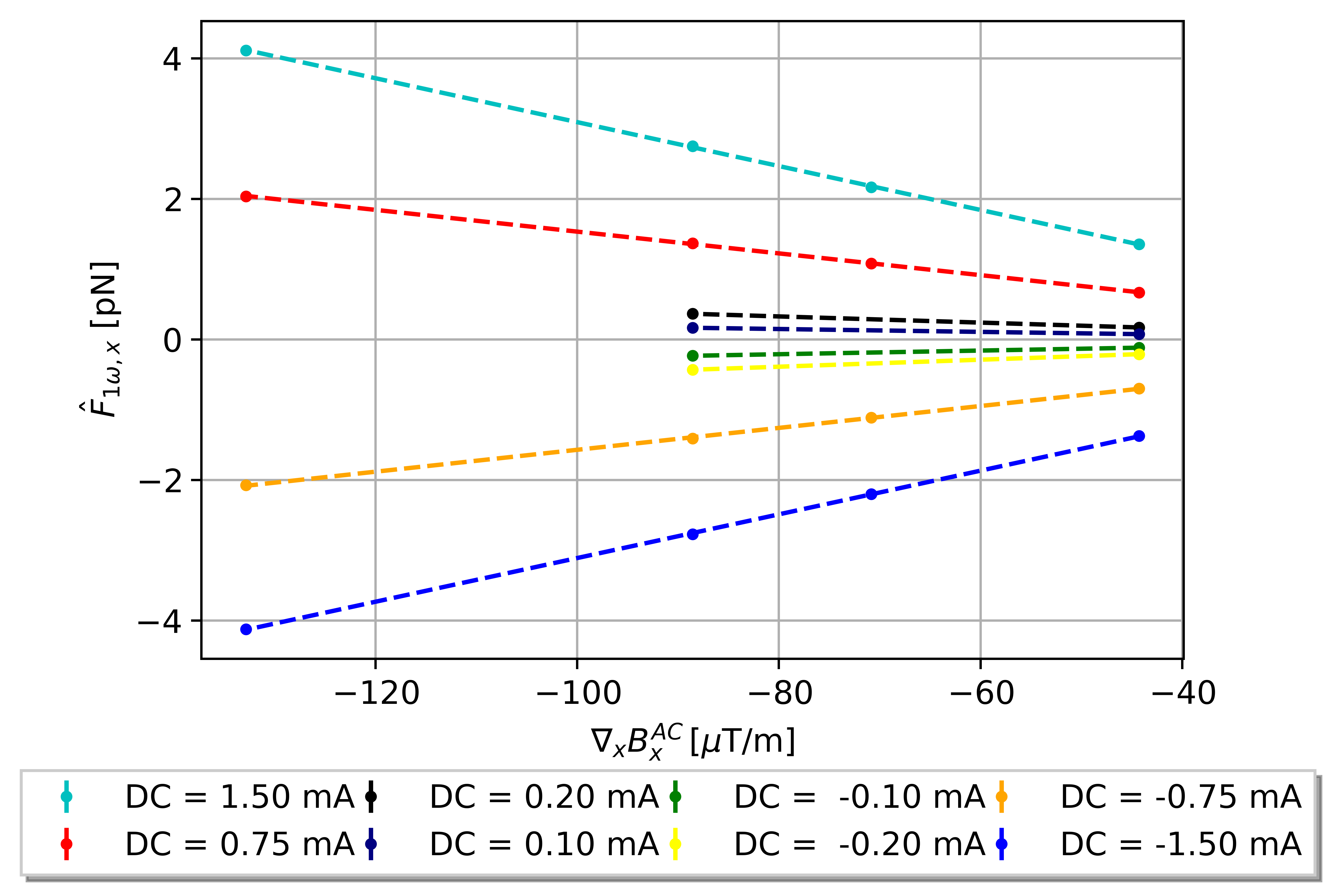}
\caption{$\hat{F}_{1\omega,x}$ on TM1 as a function of the applied AC magnetic field gradient. The different colors correspond to fixed DC values of the injected signal.}
\label{fig.F1w_fixedDCs}
\end{figure}

In order to evaluate this term, during the June $\rm 18^{th}$ run several injections with different AC field gradients were applied to the TMs. 
By doing so we can evaluate the term in brackets above at different values of the gradient $\nabla_{x} B_{x}^{AC}$. 
This is shown in Fig.~\ref{fig.F1w_fixedDCs} where we display how the $1\omega$ component of the force  changes depending on the intensity of the injected AC magnetic field gradient, for fixed DC offsets.  Each result in the plot represents an injection at 5\,mHz with AC amplitudes applied to the coil of $I^{AC} = 0.5, 0.8, 1.0, 1.5\, {\rm mA}$.

\begin{figure}[t]
\includegraphics[width=\columnwidth]{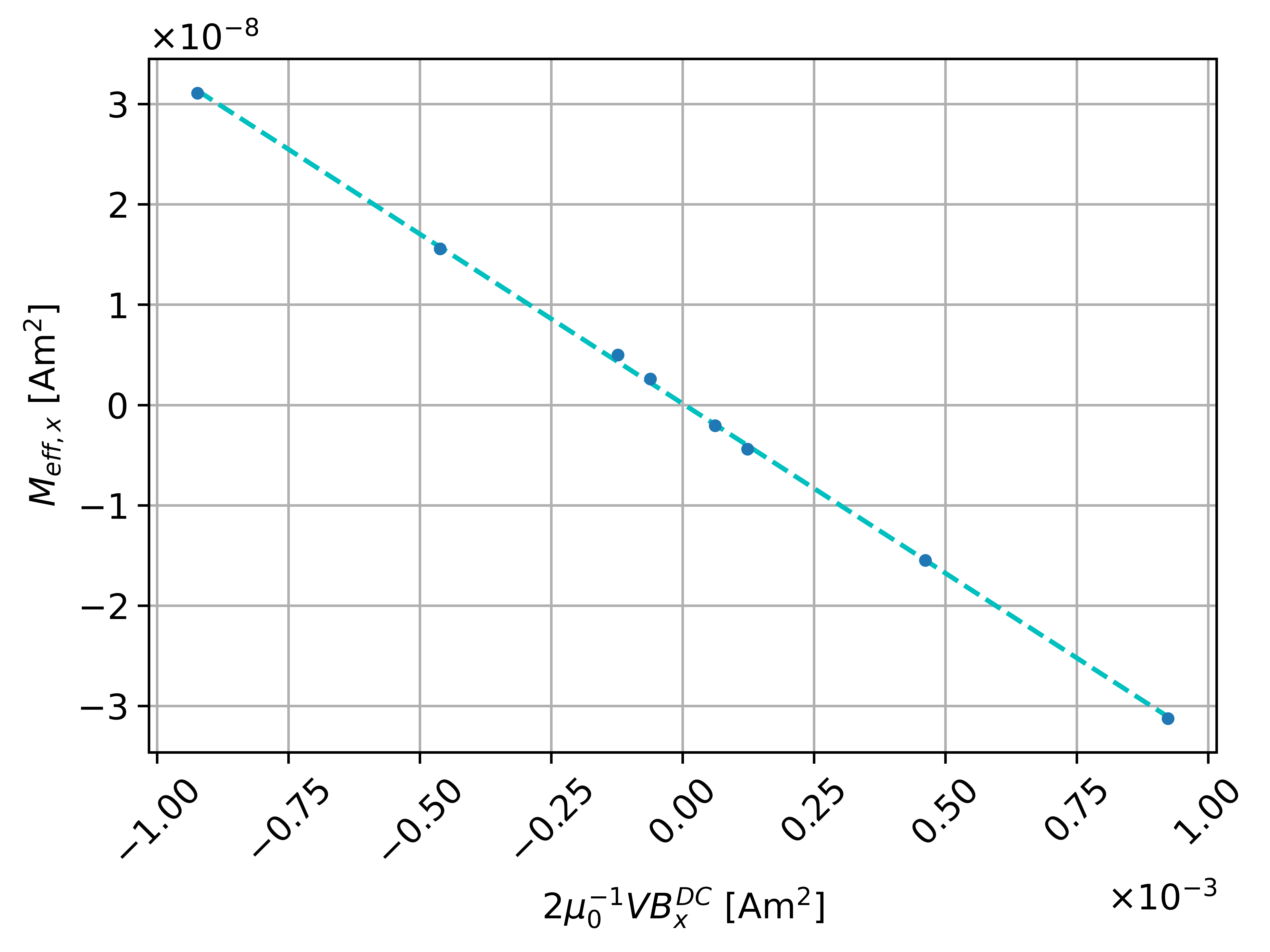}
\caption{Effective remanent magnetic moment plotted as a function of the injected DC magnetic field. The result of the fit, for an equation of the type y = mx + n, is m = $(-3.380 \pm 0.027)\times10^{-5}$ and n =  $(0.140 \pm 0.138)\times10^{-9}$ Am$^2$.}
\label{fig.Meffectives}
\end{figure}

As explained in Section \ref{sec:extraction}, by running the experiment at different DC levels we can further disentangle the dependencies of the parameters inside the brackets and  obtain the estimate for the remanent magnetic moment. Table~\ref{tbl.linesfixedDCs_coeffs} gathers the linear fits to the results that we also show in Fig.~\ref{fig.Meffectives}.
The offset parameter corresponds to the remanent magnetic moment of test mass $\#1$ in the $x$ direction, $M_{r,x} = 0.140 \pm 0.138$ nAm$^2$. 
The slope parameter is directly related to the magnetic susceptibility at $1\omega$. However, the values that were used for DC offsets of $\pm 0.1, 0.2$ mA came from the injections of April $\rm 28^{th}$ which were performed at different frequencies (3 mHz), than the rest at 5 mHz. Thus, to obtain the value of the susceptibility at 5 mHz, we will use the fit from Fig.~\ref{fig.Meffectives}, but with only the DC values of $\pm 0.75, 1.5$ mA giving a result of $\chi_{\rm{5mHz}} = (-3.3723 \pm 0.0069)\times10^{-5}$.

\begin{table}[b]
\caption{Coefficients of the fitted lines of Fig.~\ref{fig.F1w_fixedDCs} of the type $y = Ax + B$. The errors of the fits for $I^{DC} = \pm 0.1,0.2$ are 0 because there were only two points to fit the line.}
\vspace{8mm}
\begin{tabular}{ccccc}
$I^{DC} {\rm [mA]}$& A $ {\rm [nAm^{2}]}$  & B $ {\rm [fN]} $  \\  
\hline
\hline
 1.50 & $(-31.24 \pm 0.25)$ & $(-31 \pm 23)$  \\
 0.75 & $(-15.48 \pm 0.15)$ & $(-13 \pm 13)$  \\
 0.20  & $(-4.39 \pm 0 )$ & $(-23 \pm 0)$ \\
 0.10  & $(-2.04 \pm 0 )$ & $(-14 \pm 0)$  \\
 -0.10  & $(2.60 \pm 0 )$ & $(-1.3 \pm 0)$  \\
 -0.20 & $(5.00 \pm 0 )$ & $(13 \pm 0)$  \\
 -0.75 & $(15.58  \pm 0.21$ & $(-12 \pm 19)$ \\
 -1.50 & $(31.09 \pm 0.23)$ & $(-1.3 \pm 21)$ \\
\end{tabular}
\label{tbl.linesfixedDCs_coeffs}
\end{table}

Analogously, to obtain $M_{r,y}$ and $M_{r,z}$, we demodulate the amplitudes of the torque measurements around the required angles and apply Eq.~\eqref{eq.fortorques} for the respective injected magnetic fields. This way, we obtained $M_{r,y}=0.178\pm0.025$ nAm$^2$ and 
 $M_{r,z}=0.095\pm0.010$ nAm$^2$ which we will need for the background estimations. Note the large uncertainty obtained for $M_{r,x}$ compared with $M_{r,y}$ and $M_{r,z}$. The difference is due to the different methods used to extract them. The former is calculated indirectly as the offset of the linear fit of the slopes of the force at $1\omega$ while the other two components of the remanent magnetic moment are calculated directly from the measured torques along $\eta$ and $\phi$. Rotations along $\theta$ would have allowed a more precise value for $M_{r,x}$ but the interferometric system is not sensitive along such axis as it is aligned with the laser beam. The results lead to a total remanent magnetic moment of: $|\textbf{M}_r| = (0.245\pm0.081)$ nAm$^2$.

\subsection{Background magnetic field} \label{Background}

The induction of forces in the test mass by means of the controlled injection of magnetic fields allow the determination not only of the test mass magnetic parameters but also of environment parameters that contribute to the magnetic force, which is the case of the background magnetic field in the test mass position. We emphasize here that this is the way to estimate of this parameter since the magnetometers are located too far away from the test masses to guarantee a precise estimate of this variable.

\begin{figure}[t]
\includegraphics[width=\columnwidth]{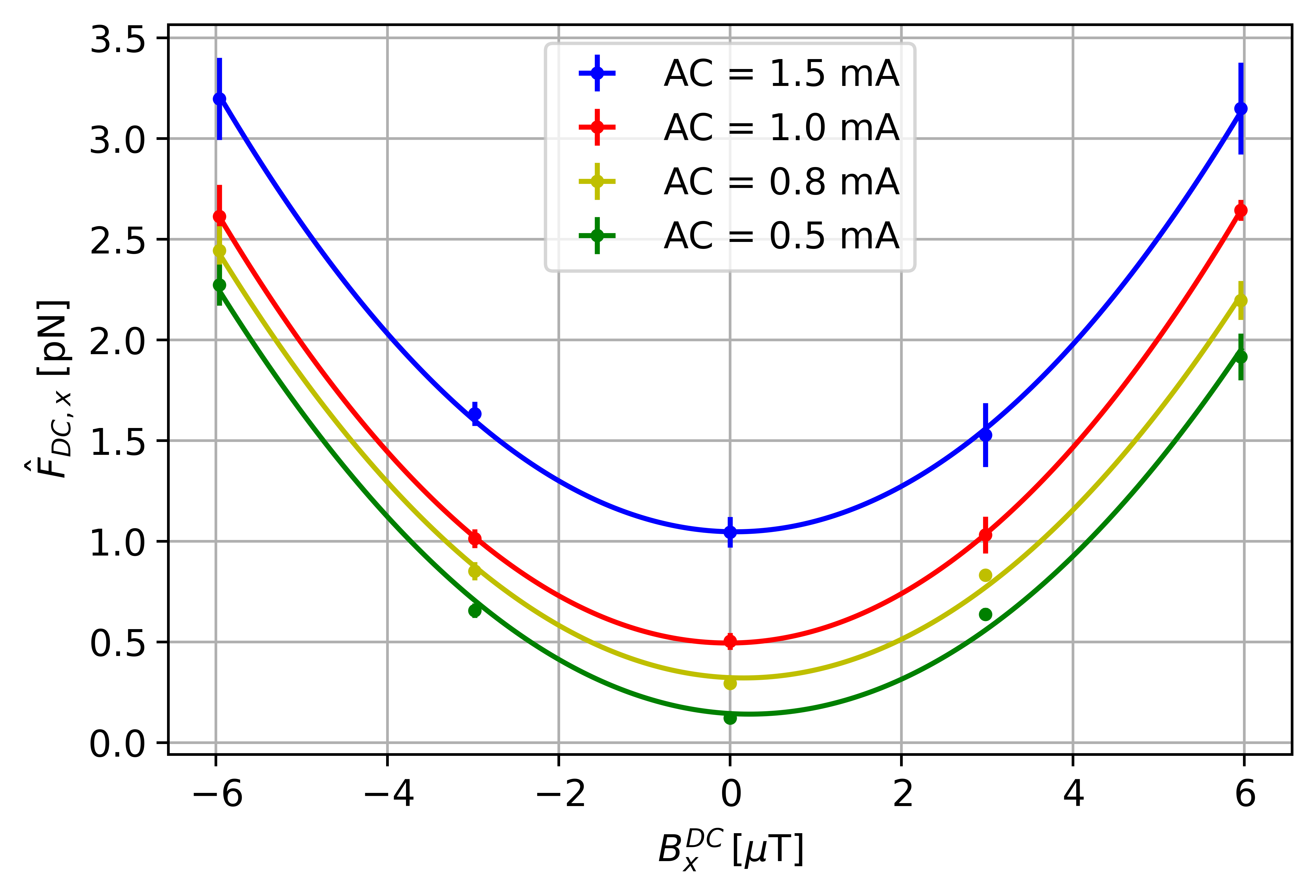}
\caption{$\hat{F}_{DC,x}$ on TM1 as a function of the injected DC magnetic field together with their respective fits to an equation of the type y = $Ax^{2} + Bx + C$. The different colors correspond to fixed AC values of the injected signal.}
\label{fig.parabolas}
\end{figure}

In order to do so we evaluate Eqs.~\eqref{eq.forceDC_DCs} using the injections of June $\rm 18^{th}$, 2016. We express the information provided by these runs by displaying the DC component of the measured force as a function of the DC component of the applied magnetic fields. As predicted, we obtain the parabolas in Fig.~\ref{fig.parabolas} for different values of the AC amplitude from where we derive the parabola coefficients of Table~\ref{tbl.parabolas_coeffs}.

\begin{table}[b]
\caption{Coefficients of the fitted parabolas of Fig.~\ref{fig.parabolas} of the type $y = A\,x^{2} + B\,x + C$.}
\vspace{8mm}
\resizebox{\columnwidth}{!}{
\begin{tabular}{ccccc}
$\rm{I}^{AC} [\rm{mA}]$& A $[\rm{N/T}^{2}](10^{-2})$ & B $[\rm{N/T}](10^{-8})$ & C $[\rm{N}](10^{-12})$ \\  
\hline
\hline
 1.5 & $(5.99 \pm 0.11)$ & $(-0.67 \pm 0.39)$ & $(1.047 \pm 0.025)$ \\
 1.0 & $(6.003 \pm 0.025)$ & $(0.265 \pm 0.089)$ & $(0.4948 \pm 0.0058)$ \\
 0.8 & $(5.64 \pm 0.16)$ & $(-1.73 \pm 0.58)$ & $(0.323 \pm 0.038)$ \\
 0.5 & $(5.50 \pm 0.22)$ & $(-2.46 \pm 0.79)$ & $(0.144 \pm 0.052)$ \\
\end{tabular}}
\label{tbl.parabolas_coeffs}
\end{table}

As derived from Eq.~\eqref{eq.forceDC_DCs}, the $A$ coefficient provides a direct estimate of the magnetic susceptibility 
\begin{equation}
A =  \left(\frac{\chi V}{\mu_0 \kappa} \right)
\label{eq:A_parabola}
\end{equation}
which, in contrast with other alternative estimates, is not dependent of the injection modulation frequency. The value obtained using Eq.~\eqref{eq:A_parabola} is $\chi_{DC}  = (-3.35 \pm 0.15)\times10^{-5}$. 

With two remaining terms, $B$ and $C$, we can build a system of equations, being the two unknowns the parameters that define the background magnetic field in the test mass position, i.e.  $B_{back., x}$ and $\nabla_{x}B_{back., x}$

\begin{equation}
\begin{split}
C =  M_{+} \nabla_{x}B_{back., x} + \frac{\chi V}{\mu_0} \Biggl[ & 3B_{back., x}\nabla_{x}B_{back., x} \nonumber \\& + \dfrac{1}{2}B_{x}^{AC}\nabla_{x}B_{x}^{AC} \Biggr],
\end{split}
\end{equation}
\begin{equation}
B  =  \frac{M_{r,x}}{\kappa} + \frac{\chi V}{\mu_0} \left[\nabla_{x} B_{back., x} + \frac{B_{back., x}}{\kappa}\right]. \label{eq.systemequations}
\end{equation}

Solving this quadratic system of equations, we obtain an expression for the background magnetic field and its gradient in the $x$ direction at the location of the TMs that we can evaluate for each of the four fit values. From the two mathematically available solutions we select the one closer to the estimates of the magnetic field and field gradient obtained during on-ground characterization of the spacecraft, which were 267 nT and -7575 nT/m, respectively~\cite{S2-ASU-TN-2523}. The values that we obtain for the in-flight estimates are $B_{back., x} = 414 \pm 74 \rm{nT}$ and $ \nabla_{x} B_{back., x} = -7400 \pm 2100 \rm{nT/m}$.

\begin{figure}[t]
\includegraphics[width=\columnwidth]{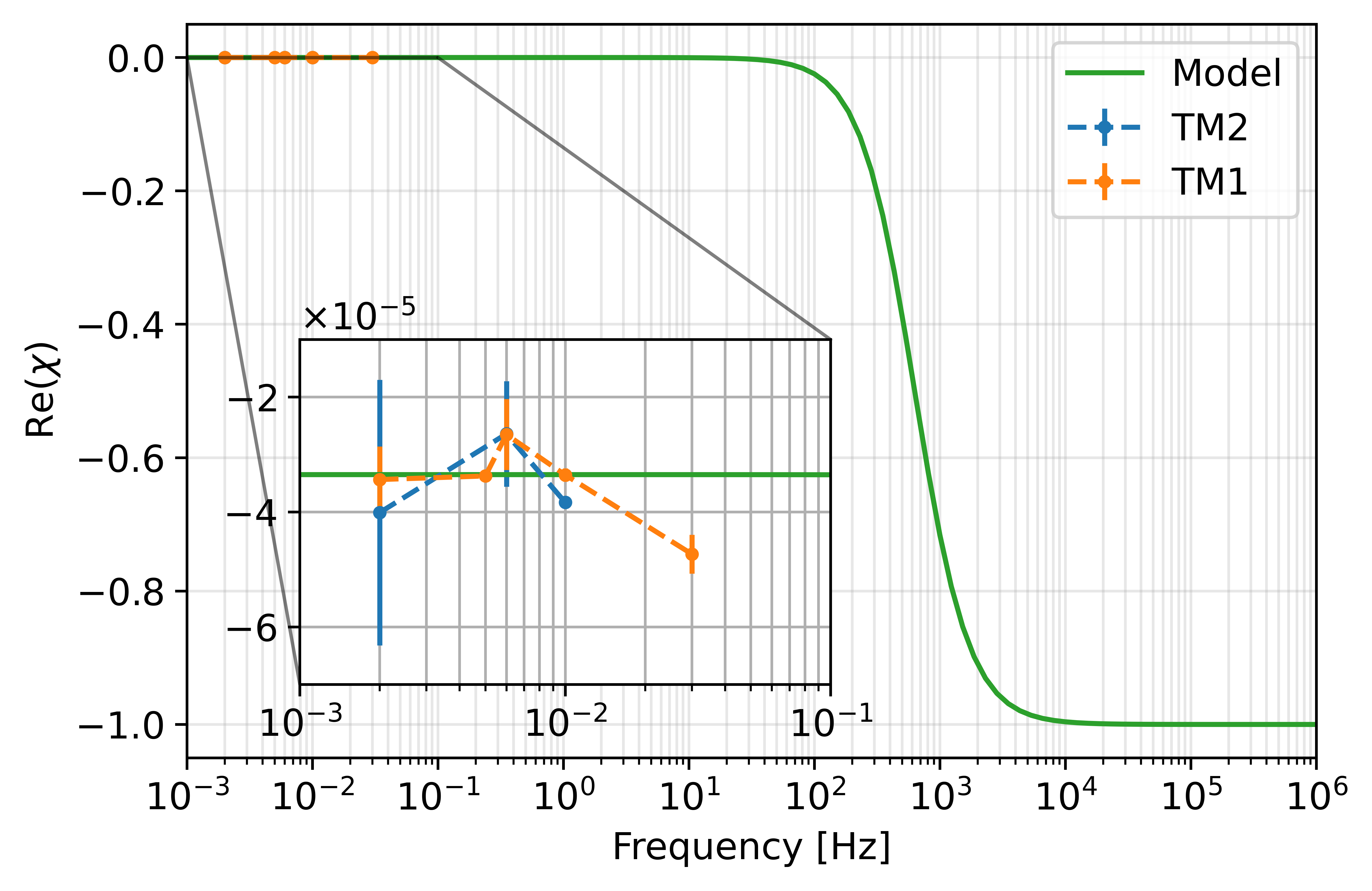}
\caption{Magnetic susceptibility of both test masses at different frequencies together with the model predicted by Eq.~\eqref{eq.chi_omega}.}
\label{fig.chi_fit_logscale}
\end{figure}

\subsection{Magnetic Susceptibility}

\begin{figure*}[t]
\centering
\includegraphics[width=0.9\textwidth]{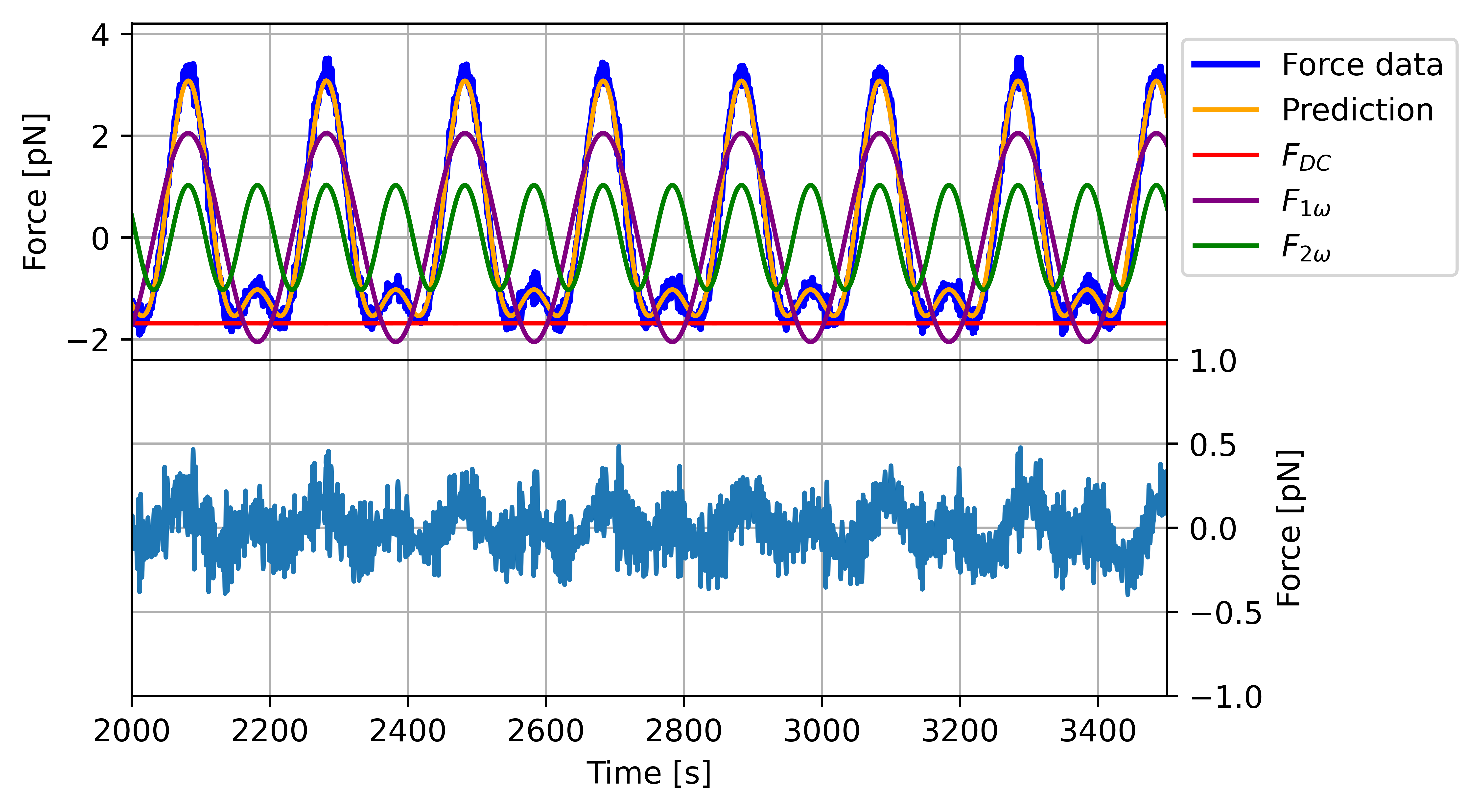}
\caption{Detrended force segment of the June injections at $I^{DC} = 0.75$ mA and $I^{AC} = 1.5$ mA compared with the predicted force model together with the three contributions at different frequencies. Below we can see the residual between the measurements and the theoretical prediction.}
\label{fig.ts_byterms}
\end{figure*} 

\begin{table}[b]
\caption{Susceptibility values of both TMs obtained at 2$\omega$ using Eq.~\eqref{eq.susc_estimate} for all the injection frequencies.}
\vspace{8mm}
\begin{tabular}{ccccc}
Frequency [mHz] & $\chi_{\rm{TM1}}$ ($10^{-5}$) & $\chi_{\rm{TM2}}$ ($10^{-5}$)    \\  
\hline
\hline
DC & $(-3.35 \pm 0.15)$ & -   \\
2 & $(-3.43 \pm 0.58)$ & $ (-4.0 \pm 2.3)$   \\
5 & $(-3.3723 \pm 0.0069)$ & -  \\
6  & $(-2.65 \pm 0.62)$ & $ (-2.64 \pm 0.92)$   \\
10 & $(-3.35 \pm 0.12)$ & $ (-3.833 \pm 0.057)$   \\
30 & $(-4.73 \pm  0.34)$ & -   \\
\end{tabular}
\label{tbl.susceptibilities}
\end{table}

We have already estimated the test mass magnetic susceptibility as a by-product of the estimate of the remanent magnetic moment and the background magnetic field and field gradient. 

The DC value of the susceptibility, that is the frequency independent part, was obtained in section \ref{Background} and the value of the susceptibility at 5 mHz was measured in section \ref{Remanent}. The rest of the measurements of the magnetic susceptibility of the TMs have been obtained using the 2$\omega$ component of the force and Eq.~\eqref{eq.susc_estimate}. These values  correspond to twice the frequency at which the injection was performed. The results of the susceptibilities can be seen in Table~\ref{tbl.susceptibilities} with the value of the frequencies at which they correspond. Using all the injections, from the three different magnetic experiments runs, the frequencies that could be calculated for the magnetic susceptibility were 2, 5, 6, 10 and 30 mHz. In coil \#2, for TM2 results, only the injections from the $\rm 29^{th}$ of April were performed, which were at frequencies 1, 3, 5 mHz meaning that only three susceptibility values could be obtained at twice their frequency.

According to~\cite{susceptibility_aprox}, the AC magnetic susceptibility of LPF TMs can be approximated  at low frequencies as
\begin{equation}
 \chi(\omega)\simeq\chi_{DC}+\frac{-i\omega\tau_e}{1+i\omega\tau_e},
 \label{eq.chi_omega}
\end{equation}
where $\chi_{DC}$ is the frequency independent term of the susceptibility and with $\tau_e$ being the magnetic susceptibility cut, i.e., the frequency at which the real and imaginary part of the magnetic susceptibility have the same value. For LPF, this value was measured on-ground to be $\tau_e=(2\pi630)^{-1}$Hz$^{-1}$~\cite{TEC-EEE/2007149/LT}.

If we now plot the measured values of the magnetic susceptibility for TMs 1 and 2 along with the curve in Eq.~\eqref{eq.chi_omega}, we obtain the plot shown in Fig.~\ref{fig.chi_fit_logscale}. We can see that all magnetic susceptibility results from Table~\ref{tbl.susceptibilities} are compatible within their uncertainty ranges as predicted from Eq.~\eqref{eq.chi_omega}, except for the one at 30 mHz. There were only three signal injected at 15 mHz, which had a low SNR and, also, showed an unexpected linear drift that had to be subtracted. We think this explains the systematics affecting the demodulation of its second harmonic at 30 mHz.  With respect to the results for TM2, we computed them for completeness but we recall that coil 2 suffered a malfunction at the beginning of operations which could explain the systematic error observed in the derived parameters.  


The results shown were obtained by heterodyne demodulation of the $\Delta g$ signal with a sinusoidal signal in-phase. However, if we made the latter be out of phase by $\pi/2$ one would expect the amplitudes measured by this method to be zero only if there were no imaginary component. Thus, the imaginary susceptibility can be obtained demodulating at $2\omega$, in quadrature. In the frequency regime that we are working in, this value is expected to be orders of magnitude smaller than the real susceptibility. The results that we obtain for the imaginary susceptibility at 10 mHz were consistent with zero, $\chi_{i} = (0.0 \pm 1.8)\times10^{-6}$. With the limits determined by the force sensitivity of the demodulation around the tenths of femtoNewtons. Thus, we can only confirm that the values of the imaginary susceptibility of the TMs are below $|\chi_{i}|< 1.8\times10^{-6}$ at 10 mHz. At the rest of frequencies the result gave a less precise upper bound for the imaginary susceptibility.

\hspace*{0.2cm}

Finally, we can evaluate the prediction of our force model from Eq.~\eqref{eq.forces} in comparison with the measured acceleration during injections by making use of the extracted magnetic parameters within this article. 
To do so, we have selected a single injection from all the ones of June $18^{\rm{th}}$, 2016 with a DC offset of 0.75 mA and an AC amplitude of 1.5 mA. We have used these values for the corresponding magnetic field DC and AC calculations together with the magnetic parameters found in the previous sections and substituted into Eqs.~\eqref{eq.3forces}. 
The results can be seen in Fig.~\ref{fig.ts_byterms} where, together with the total model force, we can see plotted the different force contributions $F_{DC}$, $F_{1\omega}$ and $F_{2\omega}$. The predicted force and the data match with a residual difference between them of $(0.0 \pm 1.9)\times10^{-13}$ N. Due to the presence of a linear drift within the data originated from the desynchronization between the clocks of the LPF measurement systems, one for the diagnostics and a different one for the optical metrology system, we are able to observe a leftovers signal within the residual data.

\section{Conclusions}\label{sec:conclusions}

The work in this paper presents the most detailed characterization of the magnetic-induced coupling on free-falling test masses in the context of gravitational wave detection in space. 
The results and method showed here were originated in the framework of the LISA Pathfinder mission. As a technology demonstrator, LPF represented a unique opportunity for an in-depth description of these effects, which can be directly transferred to LISA and future space-borne gravitational wave detectors.   

In our results, we obtain the magnetic parameters defining the response of the free-falling test masses under an external magnetic field. According to our measurements, the test mass remanent magnetic moment does not show any privileged direction, which is in agreement with the hypothesis of an isotropic, diamagnetic test mass. We have found the value of its modulus to be: $\rm |\textbf{M}_r| = (0.245\pm0.081)\,\rm{nAm}^2$, a result with much better precision than any other on-ground tests and also below the mission requirements of $\rm |\textbf{M}_r| \leq 10\,nAm^{2}$. The background magnetic field, $B_{back.,x} = 414 \pm 74$ nT, and its gradient, $\nabla_{x} B_{back.,x} = -7400 \pm 2100$ nT/m, calculated at the location of the TMs are in agreement with the values predicted by ~\cite{S2-ASU-TN-2523}. Moreover, the gradient is also below the worst case scenario predicted by ~\cite{S2-EST-TN-2026} which estimated a value of $\nabla_{x}B_{back.,x} = -11300$ nT/m. The TMs magnetic susceptibility results follow the model predicted by ~\cite{susceptibility_aprox} and they are also within the same magnitude when comparing the on-ground estimations for the DC magnetic susceptibility with values around $-2.5\times10^{-5}$ ~\cite{Bureau_International_des_Poids_et_Mesures} with our in-flight DC prediction for the magnetic susceptibility, $\chi_{DC} = (-3.35 \pm 0.15)\times10^{-5}$ and the most precisely characterized at 5 mHz of $\chi_{\rm5mHz} = (-3.3723 \pm 0.0669)\times10^{-5}$. Finally, the value of the imaginary component of the magnetic susceptibility of the TMs has been set to an upper bound at 10 mHz such that: $|\chi_{i}|< 1.8\times10^{-6}$.

These in-flight results have been acquired with precisions not achieved before with on-ground measurements. They will be of great relevance for the design of future missions with similar technologies to the ones used in LPF.

\appendix

\section{Magnetic field calculation} \label{sec:annex}

When the magnetic field is created by an induction coil, such as the case in Fig.~\ref{fig.coilsScheme}, the laws of Classical Magnetic Theory can be applied to obtain formulas which produce the values of the field and its gradient at any position in space. For slowly varying coil currents and short distances, radiative effects can be neglected. 

The system has axial symmetry, hence only parallel ($B_{x}$) and transverse ($B_{\rho}$) components of the magnetic field are different from zero. Their analytical expressions can be calculated by means of Ampère's induction laws by assuming a coil of negligible thickness and a wire winding of N turns. The result involves elliptic integrals of the first ($K(k)$) and second kind ($E(k)$). This numerical analysis does not take into account mechanical tolerances in the manufacturing and assembly; displacements and tilts induced during launch and in-orbit operations or the implicit calibration of the coil.

\subsection{Elliptic integrals}

For the calculation of the average of the magnetic field and its gradient inside the TMs volume we discretized it into 177 parts in all directions, involving the calculations of $\mathbf{B}$ and $\mathbf{\nabla B}$ in a grid of $177^{3}$ points homogeneously distributed. For each point within the grid cell of the TM, the elliptic functions $K(k)$ and $E(k)$ can be evaluated there. The equations to compute are the following, and they should be evaluated in the specified order

\begin{equation}
	\rho^{2} = y^{2} + z^{2},
\label{eq.rho}
\end{equation}
\begin{equation}
	k^{2} = \dfrac{4a\rho}{x^{2} + \left(a + \rho\right)^{2}},
\label{eq.k}
\end{equation}
\begin{equation}
	K(k) = \int_{0}^{\pi/2}{\left( 1 - k^{2}\sin^{2}\varphi \right)^{-1/2} d\varphi},
\label{eq.K}
\end{equation}
\begin{equation}
	E(k) = \int_{0}^{\pi/2}{\left( 1 - k^{2}\sin^{2}\varphi \right)^{1/2} d\varphi},
\label{eq.E}
\end{equation}

where x, y, z are the cartesian coordinates of the grid cell under calculation w.r.t. the coil center point, a is the coild radius, $\rho$ is the radial distance in cylindrical coordinates and $\varphi$ is the azimuth angle in cylindrical coordinates. 

\subsection{Magnetic field components}

The off-axis magnetic field components induced by the coil can be found to be

\begin{equation}
	B_{\rho}(x, \rho) = A_{\rho}\dfrac{x}{\rho^{3/2}}F(k),
\label{eq.Brho}
\end{equation}
\begin{equation}
	B_{x}(x, \rho) = A_{x}\rho^{-3/2}G(k) - \dfrac{\rho}{x}B_{\rho}(x, \rho),
\label{eq.Bx}
\end{equation}

where

\begin{equation}
	A_{\rho} = \dfrac{\mu_{0}}{4\pi}\dfrac{NI}{a^{1/2}},
\label{eq.Arho}
\end{equation}
\begin{equation}
	A_{x} = \dfrac{a}{2}A_{\rho},
\label{eq.Ax}
\end{equation}
\begin{equation}
	F(k) = k\left[ \dfrac{1 - k^{2}/2}{1 - k^{2}}E(k) - K(k) \right],
\label{eq.F}
\end{equation}
\begin{equation}
	G(k) = \dfrac{k^{3}}{1 - k^{2}}E(k).
\label{eq.G}
\end{equation}

From the latter equations it is easy to derive the components of the magnetic field in their cartesian components

\begin{equation}
	B_{y} = \dfrac{y}{\rho}B_{\rho} = \dfrac{y}{\sqrt{y^{2} + z^{2}}}B_{\rho},
\label{eq.By}
\end{equation}
\begin{equation}
	B_{z} = \dfrac{z}{\rho}B_{\rho} = \dfrac{z}{\sqrt{y^{2} + z^{2}}}B_{\rho}.
\label{eq.Bz}
\end{equation}

\subsection{Magnetic field gradient components}

Analytical functions for the gradients can also be calculated. Thanks to the symmetry of the system only 5 components out of all 9 possibilities need to be calculated

\begin{equation}
	\dfrac{\partial B_{\rho}}{\partial x} = A_{\rho}\rho^{-3/2}\left[ F(k) - \dfrac{x^{2}}{4a\rho}k^{3}\dfrac{F(k)}{dk} \right],
\label{eq.dBrhodx}
\end{equation}
\begin{equation}
\begin{split}
	\dfrac{\partial B_{\rho}}{\partial \rho} = A_{\rho}\dfrac{x}{\rho^{5/2}}\Biggl[& -\dfrac{3}{2}F(k) \\& + \dfrac{x^{2} + a^{2} - \rho^{2}}{8a\rho}k^{3}\dfrac{F(k)}{dk} \Biggr],
\label{eq.dBrhodrho}
\end{split}
\end{equation}
\begin{equation}
	\dfrac{\partial B_{x}}{\partial x} = -A_{x}\dfrac{x}{4a\rho^{5/2}}k^{3}\dfrac{G(k)}{dk} - \dfrac{\rho}{x}\left[ \dfrac{\partial B_{\rho}}{\partial x} - \dfrac{1}{x}B_{\rho} \right],
\label{eq.dBxdx}
\end{equation}

where

\begin{equation}
	\dfrac{F(k)}{dk} = \dfrac{1 - k^{2} + k^{4}}{\left(1 - k^{2}\right)^{2}}E(k) - \dfrac{1 - k^{2}/2}{1 - k^{2}}K(k),
\label{eq.dFdk}
\end{equation}
\begin{equation}
	\dfrac{G(k)}{dk} = \dfrac{k^{2}}{1 - k^{2}}\left[ \dfrac{4 - 2k^{2}}{1 - k^{2}}E(k) - K(k) \right].
\label{eq.dGdk}
\end{equation}

Finally, we can compute the remaining gradient components by the formulas:

\begin{equation}
	\dfrac{\partial B_{y}}{\partial x} = \dfrac{y}{\rho}\dfrac{\partial B_{\rho}}{\partial x},
\end{equation}
\begin{equation}
	\dfrac{\partial B_{z}}{\partial x} = \dfrac{z}{\rho}\dfrac{\partial B_{\rho}}{\partial x},
\end{equation}
\begin{equation}
	\dfrac{\partial B_{y}}{\partial y} = \dfrac{y^{2}}{\rho^{2}}\dfrac{\partial B_{\rho}}{\partial \rho} + \dfrac{z^{2}}{\rho^{3}}B_{\rho},
\end{equation}
\begin{equation}
	\dfrac{\partial B_{y}}{\partial z} = \dfrac{yz}{\rho^{2}}\left( \dfrac{\partial B_{\rho}}{\partial \rho} - \dfrac{1}{\rho}B_{\rho} \right).
\end{equation}
\begin{equation}
	\dfrac{\partial B_{z}}{\partial z} = \dfrac{z^{2}}{\rho^{2}}\dfrac{\partial B_{\rho}}{\partial \rho} + \dfrac{y^{2}}{\rho^{3}}B_{\rho},
\end{equation}

The averaged values obtained for a current value of 1 mA are: $\left\langle B_{x}\right\rangle = 4.465 \,\mu$T,  $\left\langle B_{y,z}\right\rangle = 0.327$ fT, $\left\langle \partial B_{x}/\partial x \right\rangle = -99.500 \,\mu$T/m, $\left\langle \partial B_{x}/\partial y,z \right\rangle = 30.119$ nT/m. Hence, the induced field can be considered to only have x components at the TMs for both the magnetic field and its gradient as the other ones are negligible. Moreover, we note that there is a linear proportionality between $\left\langle B_{x}\right\rangle$ and $\left\langle \partial B_{x}/\partial x \right\rangle$ such that $\left\langle B_{x}\right\rangle = \kappa\left\langle \partial B_{x}/\partial x \right\rangle$. This gives a value $\kappa = -0.04487$ m for the location of the TM with respect to the coil (consdering its entire volume) which is independent of the induced current through the coil and only depends on the coil dimensions and the distance to its center point. This constant factor can be determined analytically for the simpler on-axis magnetic field of a coil

\begin{equation}
	B_{x}(x) = \dfrac{\mu_{0}NIa^{2}}{2\left( x^{2} + a^{2} \right)^{3/2}},
\end{equation}
\begin{equation}
	\dfrac{\partial B_{x}(x)}{\partial x} = -\dfrac{3\mu_{0}NIa^{2}}{2\left( x^{2} + a^{2} \right)^{5/2}}x,
\end{equation}
\begin{equation}
	\kappa = \dfrac{B_{x}}{\partial B_{x}/\partial x} = -\dfrac{x^{2} + a^{2}}{3x},
\end{equation}

but is harder when involved with elliptic integrals. Thus, we simply determined its value numerically by considering different currents and distances and we obtained a numerical error many orders of magnitude below: $10^{-16}$.

\section{Magnetic torque} \label{sec:torque}

From Eq.~\eqref{eq.Torque} we can obtain the torque along the three axis of the TM: $\theta$, $\eta$ and $\phi$. However, LPF interferometric system is not sensitive to rotations around the x axis since it is the one aligned with the laser beam (thus, such given rotations do not affect the distance x between both TMs so we cannot observe them with the $\Delta g$). Hence, we will only be able to extract information from the two remaining terms $N_{\eta, \phi}$. Without loss of generality, we will analyze $N_{\eta}$ in detail as the same can be done for the other one. For simplicity, assume the induced magnetic field is only composed of an AC term such that $\mathbf{B} = \mathbf{B}^{AC}\sin(\omega t)$, then

\begin{equation}
\begin{split}
	\text{N}_{\eta} =&  \left\langle  \sin(\omega t)\left( m_{r,z}B_{x}^{AC} - m_{r,x}B_{z}^{AC}  \right) \right\rangle 
				\\& + \left\langle \left[ r_{z}\Omega_{x} - r_{x}\Omega_{z} \right] \right\rangle,
\end{split}
\label{eq.torqueEta}
\end{equation}

where

\begin{equation}
\begin{split}
	\mathbf{\Omega} =& \left( \mathbf{m}_r\cdot\mathbf{\nabla} \right)\mathbf{B}^{AC} \sin(\omega t)
				   \\& + \dfrac{\chi}{\mu_{0}}\left( \dfrac{1 - \cos(2\omega t)}{2} \right)\left( \mathbf{B}^{AC}\mathbf{\nabla} \right)\mathbf{B}^{AC},
\label{eq.Omega}
\end{split}
\end{equation}

we note that the torque can have a component at twice the injected frequency, $2\omega$. In the previous section we saw that the only non-negligible component of the induced magnetic field $\mathbf{B}^{AC}$ is in the x direction. Thus, in the first term of Eq.~\eqref{eq.torqueEta} the component proportional to $B_{z}^{AC}$ can be removed. Furthermore, we can see that the z component of $\mathbf{\Omega}$, $\Omega_{z}$, depends on $B_{z}^{AC}$ meaning that it will equal zero as well. Finally, the element that depends on $r_{z}$ cannot be cancelled so easily but note that we are integrating over the TM volume meaning that $r_{z}$ will take opposite signs due to the symmetry of the system thus, averaging to zero. The same will be true for $r_{y}$ (but not for $r_{x}$, this component disappears thanks to $\Omega_{z}$). We are left with

\begin{equation}
	\text{N}_{\eta} = \left\langle  m_{r,z}B_{x}^{AC} \right\rangle  \sin (\omega\, t).
\end{equation}

The same can be done for $N_{\phi}$

\begin{equation}
	\text{N}_{\phi} = \left\langle  -m_{r,y}B_{x}^{AC} \right\rangle  \sin (\omega\, t).
\end{equation}

This derivation also holds true if the induced magnetic field includes some background or DC components. This will result on the torque being composed of both a DC term and another one proportional to $1\omega$

\begin{equation}
\mathbf{N} = \mathbf{N}_{DC} + \mathbf{N}_{1\omega},
\label{eq:torqueTotal}
\end{equation}

where

\begin{subequations}
\label{eq.torquescomponents}
\begin{align}
\begin{split}
\mathbf{N}_{DC} = \left\langle\mathbf{m_r} \times  \mathbf{B}_0 \right\rangle,
\label{eq:torqueDC}
\end{split}
\\[0ex]
\begin{split}
\mathbf{N}_{1\omega} = \left\langle\mathbf{m_r} \times \mathbf{B}^{AC} \right\rangle  \sin (\omega\, t).
\label{eq:torque1omega}
\end{split}
\end{align}
\end{subequations}

\section{Magnetic experiments}
\label{sec.App_runs}

We list here the magnetic experiments during LISA Pathfinder operations together with the parameters used to command the coils on-board. DOY stands for day of year and f refers to the injection frequency of the current applied to the coil.

\begin{table}
\caption{1st set of injections, in coil \#1 (April 28$^{th}$, 2016).}
\begin{tabular}{ccccc}
DOY & f [mHz] & $I^{DC}$ [mA] &  $I^{AC}$ [mA] & duration [s]\\  
\hline
\hline
119  & 5 & -0.2  & 1.0 & 1000 \\
119  & 5 & -0.1  & 1.0 & 1000 \\
119  & 5 & 0.00  & 1.0 & 1000 \\
119  & 5 & +0.1  & 1.0 & 1000 \\
119  & 5 & +0.2  & 1.0 & 1000 \\
119  & 3 & -0.2  & 0.5 & 1000 \\
119  & 3 & -0.1  & 0.5 & 1000 \\
119  & 3 & 0.00  & 0.5 & 1000 \\
119  & 3 & +0.1  & 0.5 & 1000 \\
119  & 3 & +0.2  & 0.5 & 1000 \\
119  & 1 & 0.00  & 0.1 & 20000 \\
\end{tabular}
\end{table}

\begin{table}
\caption{2nd set of injections, in coil \#2 (April 29$^{th}$, 2016).}
\begin{tabular}{ccccc}
DOY & f [mHz] & $I^{DC}$ [mA] &  $I^{AC}$ [mA] & duration [s]\\  
\hline
\hline
120  & 5 & -0.2  & 1.0 & 1000 \\
120  & 5 & -0.1  & 1.0 & 1000 \\
120  & 5 & 0.00  & 1.0 & 1000 \\
120  & 5 & +0.1  & 1.0 & 1000 \\
120  & 5 & +0.2  & 1.0 & 1000 \\
120  & 3 & -0.2  & 0.5 & 1000 \\
120  & 3 & -0.1  & 0.5 & 1000 \\
120  & 3 & 0.00  & 0.5 & 1000 \\
120  & 3 & +0.1  & 0.5 & 1000 \\
120  & 3 & +0.2  & 0.5 & 1000 \\
120  & 1 & 0.00  & 0.1 & 20000 \\
\end{tabular}
\end{table}

\begin{table}
\caption{3rd set of injections, in coil \#1 (June 18$^{th}$, 2016).}
\label{tab.appendix_June}
\begin{tabular}{ccccc}
DOY & f [mHz] & $I^{DC}$ [mA] &  $I^{AC}$ [mA] & duration [s]\\  
\hline
\hline
170  & 5 & +1.5  & 1.5 & 4000 \\
170  & 5 & +1.5  & 1.0 & 4000 \\
170  & 5 & +1.5  & 0.8 & 4000 \\
170  & 5 & +1.5  & 0.5 & 4000 \\
170  & 5 & +0.75  & 1.5 & 4000 \\
170  & 5 & +0.75  & 1.0 & 4000 \\
170  & 5 & +0.75  & 0.8 & 4000 \\
170  & 5 & +0.75  & 0.5 & 4000 \\
170  & 5 & 0.00  & 1.5 & 4000 \\
170  & 5 & 0.00  & 1.0 & 4000 \\
170  & 5 & 0.00  & 0.8 & 4000 \\
170  & 5 & 0.00  & 0.5 & 4000 \\
170  & 5 & -0.75  & 1.5 & 4000 \\
170  & 5 & -0.75  & 1.0 & 4000 \\
170  & 5 & -0.75  & 0.8 & 4000 \\
170  & 5 & -0.75  & 0.5 & 4000 \\
170  & 5 & -1.5  & 1.5 & 4000 \\
170  & 5 & -1.5  & 1.0 & 4000 \\
170  & 5 & -1.5  & 0.8 & 4000 \\
170  & 5 & -1.5  & 0.5 & 4000 \\
\end{tabular}
\end{table}

\begin{table}
\caption{Last set of injections, in coil \#1 (From March 14$^{th}$, 2017 to March 16$^{th}$, 2017).}
\begin{tabular}{ccccc}
DOY & f [mHz] & $I^{DC}$ [mA] &  $I^{AC}$ [mA] & duration [s]\\  
\hline
\hline
73  & 15 & -0.5  & 0.07 & 75000 \\
74  & 15 & +0.5  & 0.07 & 75000\\
75  & 15 & +1.0  & 0.07 & 75000 \\
\end{tabular}
\end{table}






%



\newpage
\bibliographystyle{apsrev4-2}
\bibliography{library} 

\end{document}